\documentclass[%
 reprint,
superscriptaddress,
nofootinbib,
 amsmath,amssymb,
 aps,
prd,
showkeys,
]{revtex4-1}

\usepackage{graphicx}
\usepackage{dcolumn}
\usepackage{bm}
\usepackage{hyperref}
\usepackage{tabularx}
\usepackage{amsmath}
\usepackage{cleveref}
\usepackage{amssymb}
\usepackage{booktabs}
\usepackage{makecell}
\usepackage{bm}
\usepackage{multirow}
\usepackage[inline]{enumitem}
\usepackage{comment}
\usepackage{siunitx}
\usepackage{caption}
\usepackage{afterpage}
\usepackage{float}
\usepackage{hyperref}

\DeclareSIUnit[number-unit-product = ]\percent{\char`\%}
\DeclareSIUnit\foot{ft}
\DeclareSIUnit\inch{inch}
\DeclareSIUnit\pot{POT}
\DeclareSIUnit\year{years}
\DeclareSIUnit\ADC{ADC\ counts}
\DeclareSIUnit\spill{spill}
\DeclareSIUnit\ton{t}

\newcommand{\nsec}{\si{\nano\second}}
\newcommand{\psec}{\si{\pico\second}}

\begin{document}

\preprint{PRD/Particle physics experiment}

\title{Energy and Flavor Discrimination Using Precision Time Structure in On-Axis Neutrino Beams}

\author{E. Angelico} \affiliation{Enrico Fermi Institute, University of Chicago, Chicago IL 60637}
\author{J. Eisch}\affiliation{Iowa State University, Ames IA 50011}
\author{A. Elagin}\affiliation{Enrico Fermi Institute, University of Chicago, Chicago IL 60637}
\author{H. J. Frisch}\affiliation{Enrico Fermi Institute, University of Chicago, Chicago IL 60637}
\author{S. Nagaitsev} \affiliation{Fermi National Accelerator Laboratory, Batavia IL 60510}\affiliation{Enrico Fermi Institute, University of Chicago, Chicago IL 60637}
\author{M. Wetstein}\affiliation{Iowa State University, Ames IA 50011}

\date{\today}

\begin{abstract}
We propose to use a higher-frequency RF bunch structure for the primary proton beam on target and precision timing to select different energy and flavor spectra from a wide-band neutrino beam, based on the relative arrival times of the neutrinos with respect to the RF bunch structure. This `stroboscopic' approach is complementary to techniques that select different neutrino energy spectra based on the angle with respect to the beam axis. A timing-based approach allows for the selection of varying energy spectra from the same on-axis detector, and applies equally to both the near and far detectors in an oscillation experiment. Energy and flavor discrimination of neutrinos produced by hadrons in-flight will require proton bunch lengths on the order of 100~\psec{} and commensurate time resolution in the detector. Correlating neutrino events with the parent proton interaction is currently limited by the nanosecond-scale width of the proton bunches impinging on the target. We show that these limitations can be addressed by using a superconducting RF cavity to rebunch the present 53.1 MHz RF bunch structure with a factor of 10 higher RF frequency, thus attaining the required shorter bunch length.  
\end{abstract}

\keywords{Timing, Neutrino, Oscillation, RF-Structure, Proton Bunch}
\maketitle


\section{Introduction}
\label{sec:intro}
A deeper understanding of the neutrino sector, including CP violation, the mass hierarchy, and deviations from unitarity of the Pontecorvo-Maki-Nakagawa-Sakata (PMNS) matrix, hinges on high precision, increasingly systematics-dominated measurements of neutrino oscillation parameters from neutrino beams~\cite{annrev,PhysRevD.96.095018}. Neutrinos from accelerator beams span a wide range of energies due to the kinematics of the decays and limitations of magnetic focusing of the parent hadrons. This has the desirable effect of enabling measurement of the energy-dependent oscillation pattern at multiple neutrino energies. However, it is also the source of two limiting systematics: uncertainty on the underlying neutrino flux and uncertainty in energy reconstruction~\cite{Benhar2017}. 


One technique for addressing these uncertainties is to select a narrower energy spectrum by locating the detectors off-axis from the pointing of the beam, a technique notably exploited by the NOvA~\cite{NOvAtdr} and T2K~\cite{T2Kconcept} experiments. However, off-axis detection, such as that recently proposed for the near detectors in T2K~\cite{T2Kconcept} and DUNE~\cite{Nuprism, DUNEprism}, comes at the cost of reduced flux and restricted beam energy. These prismatic approaches look at the full flux and sample multiple off-axis angles in the same detector, but their application is limited to near detectors.  


A complementary method for selecting different energy spectra within a neutrino beam that applies to both near and far detectors exploits the differing velocities of the parent hadrons. Lower energy pions and kaons travel more slowly, especially as they approach sub-relativistic energies. As a consequence, lower energy neutrinos are created and arrive at the detector later than higher energy neutrinos created by the same proton bunch. Selecting later-arriving neutrinos would provide an increasingly pure low-energy subset of the overall flux.\footnote{As shown below, the time difference from hadron travel outweighs the compensating effect that higher energy hadrons live longer.}

In addition to selecting different energy spectra from a wide-band neutrino beam, timing can be used to isolate lepton flavor and parent-hadron components of the neutrino flux, as well as to search for prompt production of new particles, such as light neutral leptons~\cite{shrock} or light dark matter~\cite{DUNEprismLDM, deGouvea2019}.

The idea of using neutrino arrival times to resolve kinematic details has a long
history. In 1998 M. Goldhaber pointed out that neutrinos from SN1987A were detected
earlier in Kamiokande than in IMB due to the correlation of energy and time of production~\cite{mgoldhaber}. The energy-dependent time
evolution of supernova neutrinos has also been proposed as a means of determining the mass hierarchy~\cite{Supernova_time_hierarchy}. In the context of neutrino beams, several notable efforts have utilized bunch timing to place limits on neutrino velocity~\cite{OPERAspeed, MINOSspeed, T2Kspeed}. MiniBooNE has recently published several analyses exploiting timing to select stopped kaons from the NuMI beam~\cite{MiniBooNEstoppedK} and to search for heavy dark matter particles~\cite{MiniBooNEdm}.

The MiniBooNE collaboration has also explored the idea of using timing relative to the RF structure of the proton beam  to select on the neutrino energy spectrum. However, efforts to select different energy spectra on the basis of beam timing have been limited due to two considerations: (1) the $\sim$1~\nsec{} spread of the proton bunch impinging on the target washes out most of the effect and (2) the time resolutions of the detectors themselves were insufficient to see the O(100)~\psec{} effect. 
 
\begin{figure*}[!htb]
\begin{center}
\includegraphics[width=0.9\textwidth]{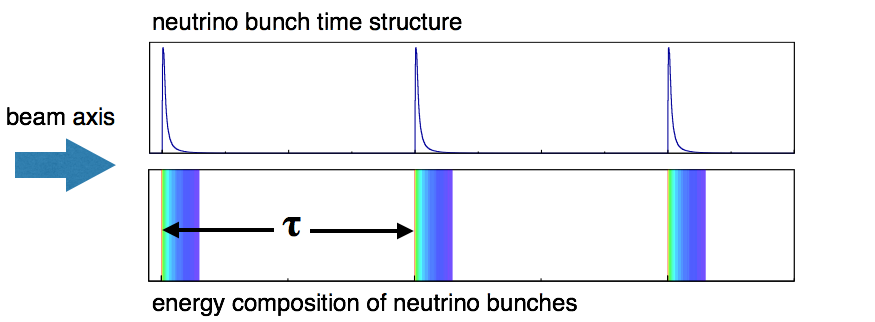}
\caption{A `snap-shot' of the neutrino wave-front moving through a neutrino detector. The top plot (blue curve) is the flux of neutrinos at a downstream detetor as a function of time. The period of the RF bunches is $\tau$; higher energy neutrinos (red - yellow) arrive before lower energy (light blue - dark blue)}\label{fig:wave_fronts}
\end{center}
\end{figure*}


Here we revisit the idea of using beam timing to select different energy components of the neutrino flux with a higher-frequency RF structure superimposed on the proton beam after normal acceleration but before extraction to make shorter proton bunches.  We are proposing to take advantage of a well-developed superconducting RF cavity technology for electron storage rings, the 500 MHz Cornell B-cell cavity~\cite{Intro_srf}. A single such cavity would be needed to implement this proposal.  Additional modest developments would be needed to re-design the RF frequency for 531 MHz and to address potential issues associated with a factor of 2.5 higher beam currents.

An overview of the `stroboscopic' approach is presented in Section~\ref{approach}, including a detailed discussion of the mechanism of energy and flavor separation by `time slicing' relative to the parent proton bunch structure, and the impact that non-zero proton bunch size has on energy separation.  Section~\ref{Fermilab} describes
the outlines of a proposed first implementation at Fermilab.
Section~\ref{RF} explores a strategy for rebunching 120 GeV protons from the Fermilab Main Injector at a higher RF frequency to achieve 150-300~\psec{} longitudinal profiles. Bunch profiles are presented from simulations of a 531 MHz cavity whose applied voltage is ramped up while the 53.1 MHz RF cavities are ramped down. Section~\ref{results} summarizes the energy spectra for electron and muon neutrinos and anti-neutrinos for the 531 MHz beam profiles of Section~\ref{RF}. Conclusions are presented in Section~\ref{conclusions}.

%
%
\section{The Stroboscopic Approach}
\label{approach}
We have dubbed the time-slicing of neutrino events relative to the
time of their parent proton bunch a `stroboscopic' approach, as it is
essentially successive snapshots of the neutrino bunch with
different energy spectra and different neutrino family populations in
each time bin. We describe the basic ingredients of the technique
below.

\subsection{Neutrino bunch time structure at the target and detectors}
\label{wave_front}

The RF structure of the proton beam at the target is imprinted into the time structure of the parent hadrons, and thus also the time structure of the produced neutrinos, as illustrated schematically in Figure~\ref{fig:wave_fronts}. Neutrinos from lower energy hadron parents (indicated in blue) tend to arrive later than neutrinos from higher energy parents (indicated in red), as described below in Section~\ref{mechanism}.  The spacing of the resulting neutrino bunches is given by the period of the proton RF structure, $\tau$. The length of the neutrino bunch depends on both the proton bunch length and the subsequent decay of hadrons that produce neutrinos. An additional factor is the different path lengths of the parent hadrons through the neutrino focusing horn system and decay region. 

\begin{figure}[t]
	\centering
        \includegraphics[width=8.6cm]{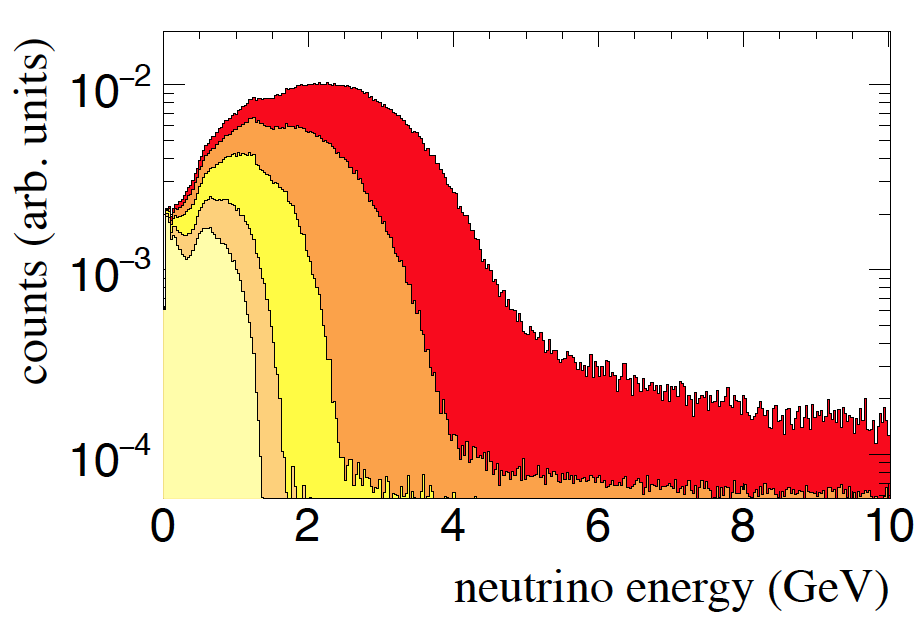} 
\caption{The simulated DUNE forward horn current flux (red), with the
          fluxes corresponding to increasingly later time-cuts on the
          bunch time, assuming no time spread of the protons on
          target: keeping all neutrinos that arrive 250~\psec{} after the start of the neutrino bunch
          (orange), 500~\psec{} after (yellow), 750~\psec{} (dark beige), 1~\nsec{} (light beige).} \label{fig:timeslices1}
\end{figure}

Time-slicing relative to the arrival of the proton bunch produces different neutrino energy spectra in each time-slice. 
To illustrate this approach,  Fig~\ref{fig:timeslices1} shows the simulated beam energy spectrum for different selection cuts with respect to the start time of the neutrino bunch in Forward Horn Current (FHC) operation of the Long-Baseline Neutrino Facility (LBNF) beam~\cite{Alion:2016uaj}. 

To discriminate neutrino interactions based on relative arrival times, one must have data on the time of arrival of the proton bunch at the target. After production, each neutrino travels for a time approximately equal to $L/c$, where $c$ is the speed of light and $L$ is the distance to the detector. The neutrinos from a given proton bunch arrive at the detector before a signal can be directly transmitted electromagnetically. Consequently, each bunch needs to be time-stamped locally at the accelerator complex and that time stamp must be synchronized with the detector at sufficient resolution. 

The spectra will also depend on the family type of the neutrino, as electrons and muons are produced in different ratios from pion and kaon decays, with electrons being highly suppressed in pion decays. Tau neutrinos, while rare, will be predominantly produced by short-lived parents, and will be enhanced in the first time slice.

\subsection{Derivation of the Effect}
\label{mechanism}

The difference in arrival time of a neutrino from a sub-relativistic
parent hadron of energy $E$ with respect to a high energy hadron traveling with
speed $\sim c$ is given by:

\begin{align}
\Delta t(E) &= \tau (E) [1 - \beta (E)]
\end{align}
where $\tau (E)$ is the lifetime of the lower energy hadron in the lab
frame. The time spreading will only occur until the decay of the sub-relativistic hadrons, at which point the daughter neutrinos will propagate at c.

Rewriting in terms of the hadron lifetime in the rest frame, $\tau_0$, we get~\footnote{We use natural units, in which mass is defined as $m=mc^2$ and momentum is defined as $p=pc$, so that energy, mass, and momentum can all be quoted in GeV.}:

\begin{equation}
\Delta t(E) = \left(\gamma  \tau_0\right) \left[1 - \sqrt{ 1 - 1/\gamma  ^2 }\right] 
\label{eq:beamkinematicsandtiming}
\end{equation}
where $\gamma  = \frac{E}{m}$. Taking the limiting approximations where $\gamma  \gg 1$ and $\gamma  \simeq  1$:

\begin{align}
&\Delta t(\gamma  \gg 1) \simeq \frac{\tau_0}{2\gamma} \\
&\Delta t(\gamma  \simeq 1) \simeq \tau _0
\end{align}
As one would expect, at the lowest energies, where the hadron is
essentially at rest in the lab frame, $\Delta t(E)$ approaches the
lifetime of the hadron in the rest frame, $\tau_0$. At high energies,
the speed of the lower-energy hadron approaches that of the higher
energy hadron and thus $\Delta t(E)$ goes to zero.

\begin{figure*}[!htb]
\centering
 \includegraphics[width=0.9\textwidth]{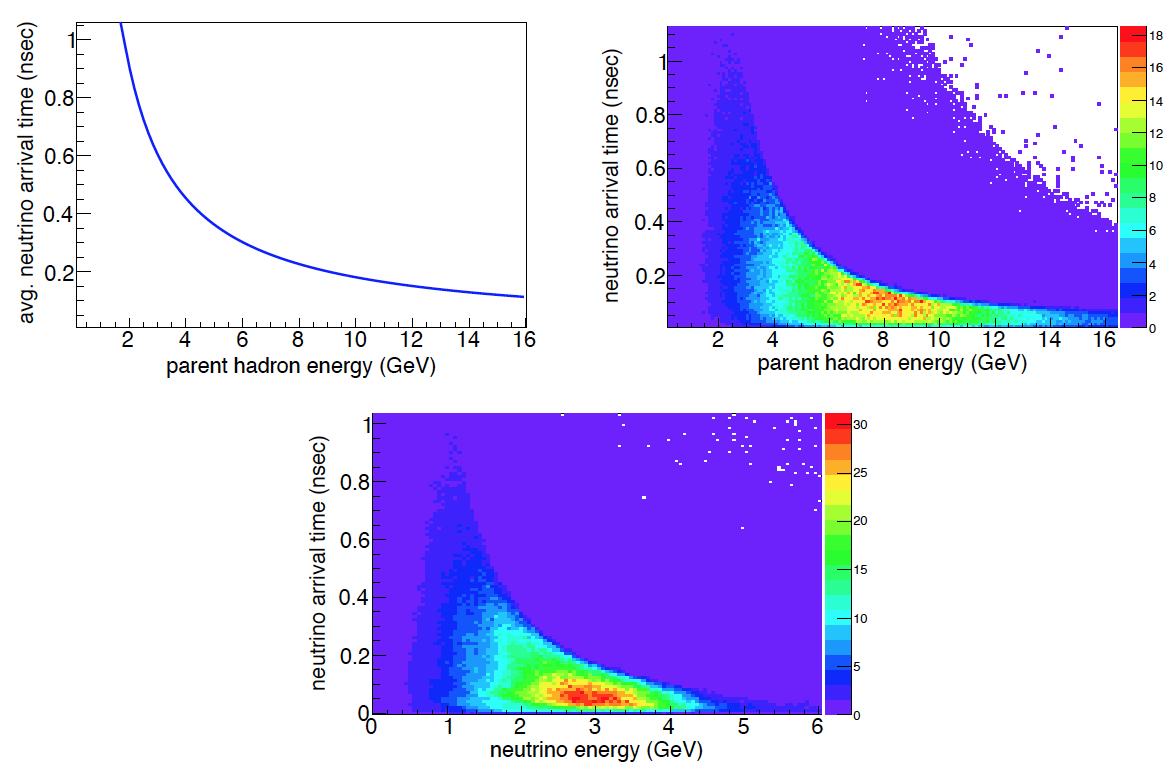} 
  
	\caption{Top left: A plot of mean relative neutrino arrival time as a function of the parent pion energy as derived from Eq~\ref{eq:beamkinematicsandtiming}. Top right: A full simulation of the relative neutrino arrival times versus parent hadron energies. The color scale represents intensity in linear arbitrary units. Bottom: The equivalent distribution of relative neutrino arrival times versus neutrino energy.}
		\label{fig:beamkinematicsandtiming}
\end{figure*}

Figure~\ref{fig:beamkinematicsandtiming} shows a graph of Equation~\ref{eq:beamkinematicsandtiming}, alongside plots of the simulated distributions between $\Delta t(E)$ and $E$ for the DUNE flux ~\cite{Alion:2016uaj} for an idealized case where all protons strike the target simultaneously (zero bunch width).

\subsection{Achieving Sufficiently Short Proton Bunch Size}
\label{protonbunchsize}

\begin{figure}[!htb]
	\begin{center}
        \includegraphics[width=8.6cm]{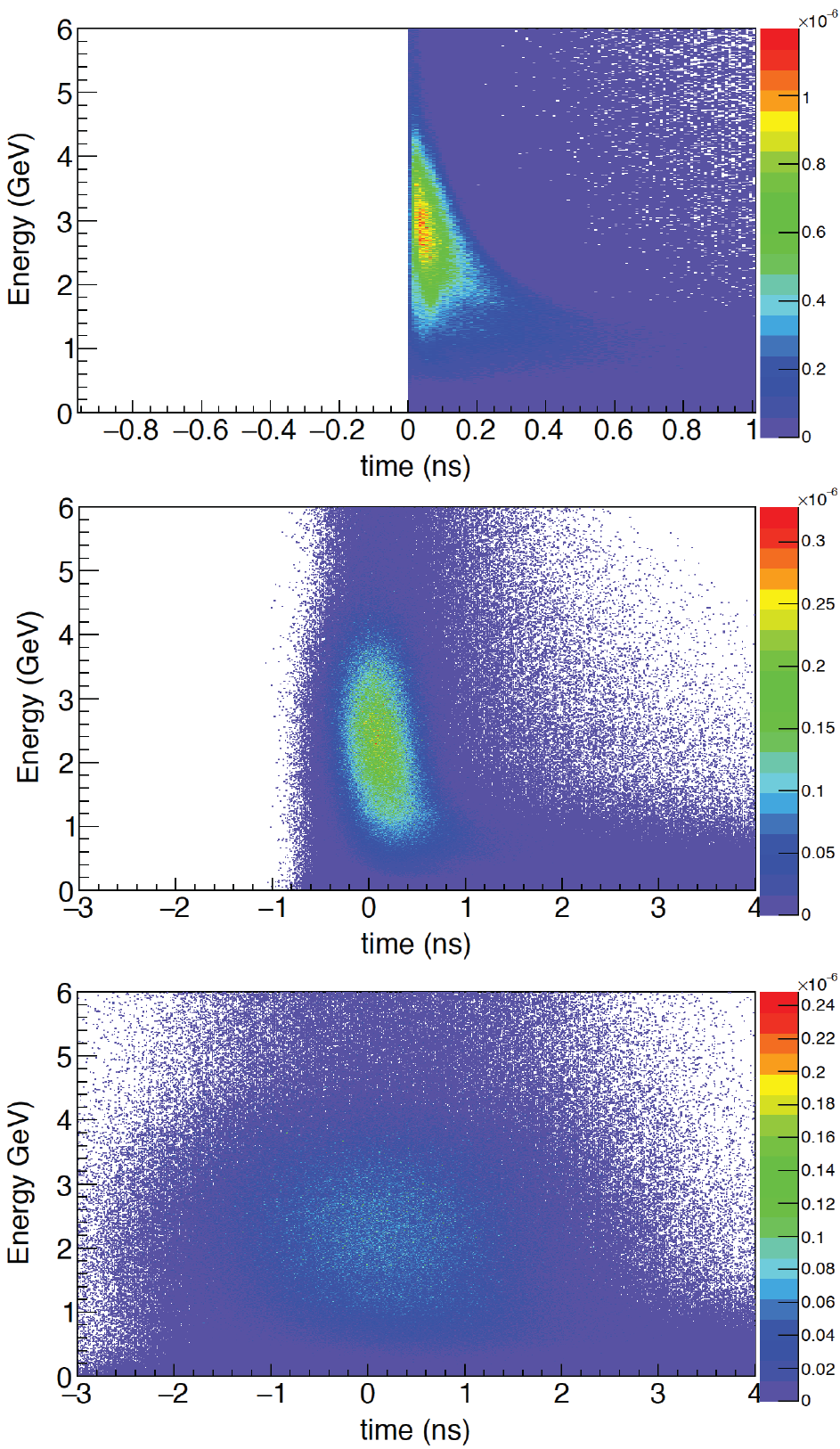}
	\end{center}
	\caption{Neutrino energy plotted against the arrival time of the neutrinos relative to proton bunch center, for three scenarios. The color scale is intensity in linear arbitrary units. Top: An idealized case where all protons strike the target simultaneously. Center: A scenario where the incident proton bunch has an RMS of 250~\psec{}. Bottom: The case where the proton incident bunch has an RMS of 1~\nsec{}, consistent with the nominal RF structure of the LBNF beam. }
		\label{fig:TimeVsE}
\end{figure}

The differentiation of pion energies based on the neutrino arrival time spectrum is tightly peaked in time, with a broad tail. This timing effect is smeared out by the time spread of the proton bunch impinging on the target. Figure~\ref{fig:TimeVsE} shows how the energy and time correlation of the neutrinos becomes smeared when the proton bunch has a non-zero time-width. At the roughly 1~\nsec{} bunch lengths typical of the Fermilab neutrino facilities the correlation between timing and energy is mostly washed out except for a small, low-statistics tail.

Fully exploiting the stroboscopic technique would require much shorter bunch sizes, approaching 100~\psec{}. These shorter bunch times can be achieved by rebunching the proton beam at a higher frequency.

We present a scheme for rebunching the Fermilab Main Injector (MI) proton beam, superimposing a higher 10th harmonic frequency substructure (see Section \ref{RF}). With existing RF technology one could adiabatically change the beam from $\sim$1~\nsec{} long bunches spaced every $\sim$20~\nsec{} into $\sim$100~\psec{} bunches spaced by $\sim$2~\nsec{}. We refer to these bunches as `531 MHz bunches'. These bunches would be both adequately short and separated to capture the desired energy and flavor effects. In the following sections, we outline the technical steps and hardware needed to rebunch the proton beam.


%
%
\section{Modification of the Current Fermilab Neutrino Facilities}
\label{Fermilab}
The proposed `stroboscopic' scheme requires one additional superconducting RF cavity to the current neutrino production facilities in the Fermilab Main Injector. This cavity will be at a 10th harmonic of the current 53.1 MHz RF cavities to superimpose a hyper-fine bunch structure on the protons. The acceleration procedure will be unchanged, followed by a ramp-down of the existing 53.1 MHz cavities and a ramp-up of the  10th harmonic cavity (531 MHz). Monitoring systems with precise time resolution will tag the interaction time of the 531 MHz bunches at the target. Precision timing information will be needed at both the neutrino detector and the proton target to synchronize neutrino events to RF bunches.  Bunch profile synchronization will require data acquisition (DAQ) infrastructure to hold and match the two data streams from the target and detector.

%

\begin{figure}[!h]
	\centering
           	\includegraphics[width=8.6cm]{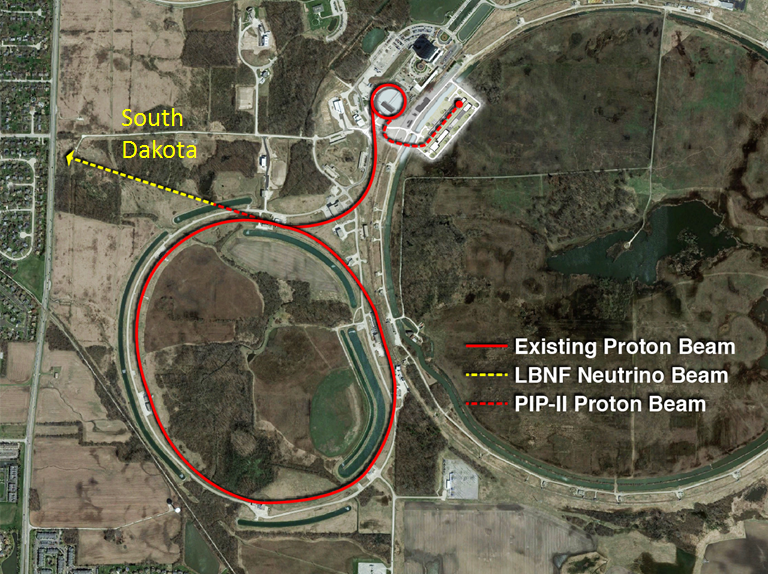}
           	 
	\caption{An aerial view of the Fermilab Accelerator complex, including the proposed PIP-II linac (dashed red line), the Booster and the Main Injector (solid red lines), and the LBNF extraction line (dashed yellow line). (Credit: Fermilab)}
	\label{fig:fermilab_facility_aerial}
\end{figure}

Figure~\ref{fig:fermilab_facility_aerial} shows an aerial view of the Fermilab Accelerator complex. The Main Injector (MI) (red oval in the Figure) accelerates protons from their injection energy of 8 GeV to 120 GeV, after which they are extracted and directed onto the neutrino target (yellow dashed line). The accelerating frequency at flat-top is 53.1 MHz. The typical MI cycle time~\cite{vaia} is 1.2 s, with a typical spill length of 11 $\mu$s. The proposed rebunching process would add no more than 60 ms to the present MI cycle. 

The proposed superposition of a higher-frequency RF on the current 53 MHz of the Main Injector is subject to several operational constraints. First, the aperture of the 531 MHz RF cavity must be large enough not to limit the MI aperture. Second, the 531 MHz cavity will operate at a constant frequency since this rebunching is performed only at the flat-top energy. Finally, a relatively high RF voltage ($\sim$2 MV) needed for rebunching, lends itself well to a single superconducting RF cavity.  A loss of integrated protons-on-target (POT) due to the increase in MI cycle time to accommodate the RF transition should be small; we take 5\% as a nominal limit. Lastly, cost and schedule indicate finding an existing hardware solution that can be purchased, with modest additional development.


\section{Accelerator Modifications: Rebunching protons at a higher RF frequency}
\label{RF}

Here we give an example implementation of creating bunches with longitudinal widths on the order of 200~\psec{} in the Main Injector (MI).

\subsection{Properties of the present Main Injector proton bunch}
\begin{figure*}[!htb]
	\centering
        \includegraphics[width=0.75\textwidth]{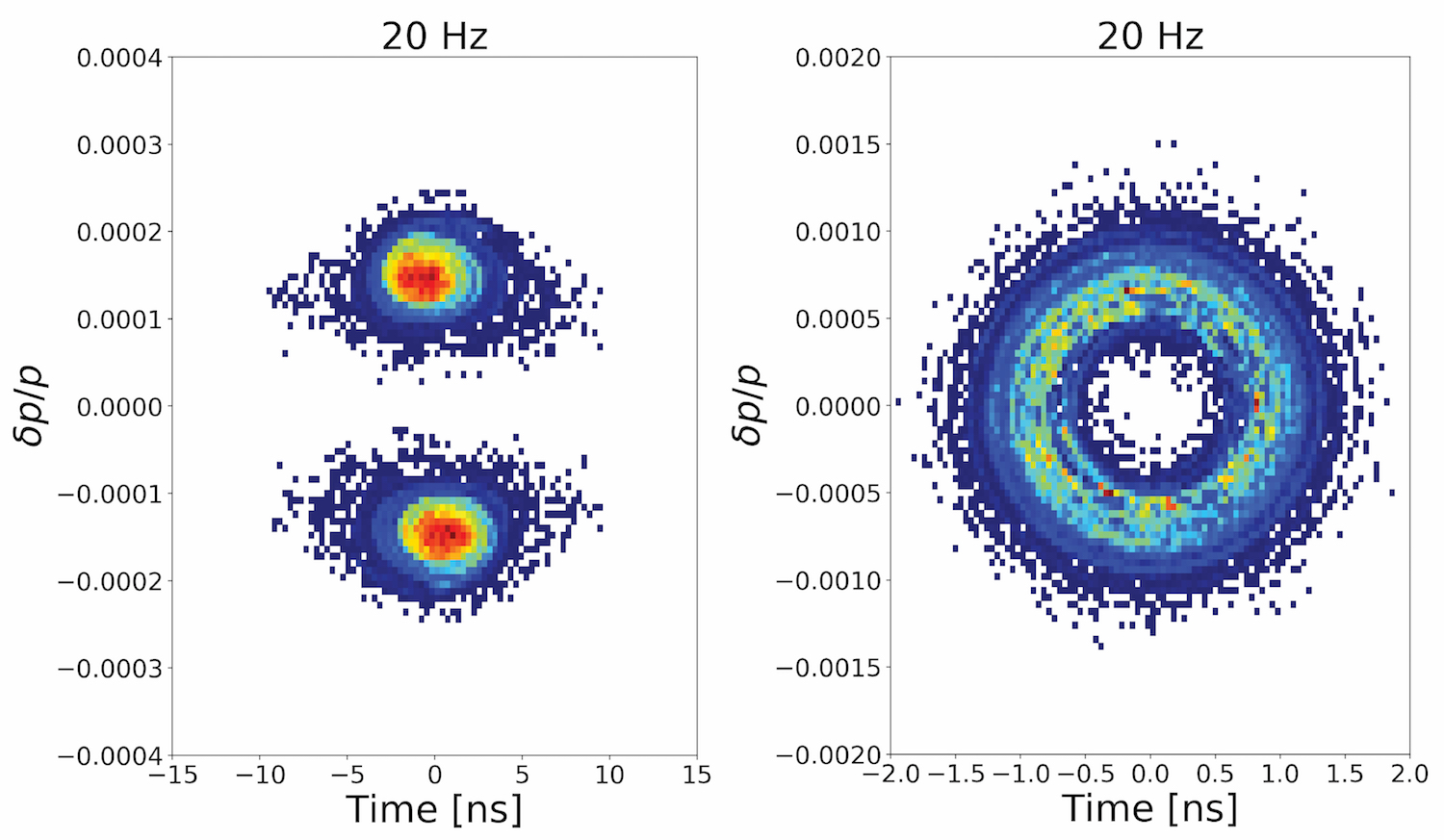}
	\caption{Left: A simulation of two Booster bunches during slip-stacking for 20 Hz booster injection (future scenario). Right: A simulation of MI bunches after acceleration to 120 GeV for 20 Hz booster injection. Color represents density (red is highest density, blue is lowest). The RF voltage amplitude is 4.6 MV. }\label{fig:MI_distributions}
\end{figure*}

The Fermilab Main Injector~\cite{Convery:2018} is filled with protons in a single turn from the Recycler ring.  The Recycler serves as an accumulator ring for batches of 8 GeV/c protons from the Booster synchrotron, operating at 15 Hz at present and at 20 Hz in the future for the LBNF. Each batch contains 84 RF buckets at 53.1 MHz. The 3.3 km long Recycler circumference is exactly 7 times the Booster circumference, or 588 buckets, but only 6 of the 7 are filled with beam, leaving an abort and extraction gap of 84 buckets that gives time for the kicker to direct the beam into the NuMI target. During every cycle of 1.2 s, the 6+6 Booster batches containing $\sim10^{14}$ protons are slip-stacked at 8 GeV/c in the Recycler by injecting them at a slightly different energy~\cite{slipstack}. When they line-up longitudinally, are captured by the 53 MHz RF system. The left side of Figure~\ref{fig:MI_distributions} shows two Booster bunches at the final moment of slip-stacking, just before being captured by the Recycler RF system in a single RF bucket~\cite{Ainsworth:IPAC2017-WEPVA039}. 

The Recycler beam is then transferred to the Main Injector, accelerated to 120 GeV and extracted to the neutrino production target downstream, creating 9.5 $\mu$s of bunched neutrino beam~\cite{PhysRevSTAB.16.071001}.

In the theory of synchrotron motion in a circular ring, the number of RF buckets in the ring is called the harmonic number, $h = f_{rf}/{f_0}$ (=588 for the Main Injector), where $f_0$ is the revolution frequency around the entire ring and $f_{rf}$ is the frequency of the RF accelerating cavities. The longitudinal position of a particle $z_i$ in any bunch can be written as an RF phase angle that is periodic with the circumference: $z_i = \frac{c}{hf_0}\phi_i = \frac{c}{f_{rf}} \phi_i$. The longitudinal phase can also be translated into a longitudinal time coordinate in the bunch $t_i = \frac{1}{f_{rf}}\frac{\phi_i}{2\pi}$. 

The RF phase, $\phi_i$, of an individual proton in the bunch changes depending on its momentum. Each proton in the bunch occupies a point in the
phase space $(\phi_i, \delta_i)$, where $\delta_i = (p_i - p_0)/p_0 = \delta p_i/p_0$ is the momentum fraction relative to a synchronous particle momentum (beam design momentum). A particle with $\delta = 0$ will not change its longitudinal position in the bunch over time. 

Each MI RF period is about 20~\nsec{} long but the protons only occupy a small fraction of the bucket, about  1~\nsec{} rms. Proton bunches from the Booster are extracted to the Recycler with a longitudinal phase-space area, $A$, of about 0.15 eV\,s (95\% of particles). Each MI bunch is a combination of two slip-stacked Booster bunches, which leads to about 0.7 to 1 eV\,s of phase space area after acceleration to 120 GeV (95\% of particles)~\cite{PhysRevSTAB.16.071001}. The right side of Figure~\ref{fig:MI_distributions} shows the MI bunches at 120 GeV, prior to extraction.

At 1 eV\,s, this implies that the relative rms momentum spread is
\begin{equation}
\delta p_{\mathrm{rms}} = \frac{A}{6\pi t_{\mathrm{rms}}} \simeq \frac{1 \, [\mathrm{eV\,s}]}{6\pi\times 1\, [\mathrm{ns}]}  \simeq 53 \, \mathrm{MeV/c}
\end{equation}
or a $\delta$ of $4.4\times 10^{-4}$. For the purposes of the study described in Section~\ref{simulation} below, we generated an initial particle distribution using two overlaid normal distributions with 95\% of the protons in one bunch occupying an area of 1 eV\,s; the resulting initial distribution of protons is shown in red in Figure \ref{fig:bunch_distributions}.

\subsection{Higher frequency RF structure and hardware}

We propose to rebunch the Main Injector proton beam using a 10-times higher RF frequency ($h=5880$) for stroboscopic neutrino physics. 

At present, commercially-available single-cell superconducting RF cavities at 500 MHz are in use in electron storage rings like CESR and the Canadian Light Source \cite{Belom:SRF2007}. These cavities are designed to operate at a 2 MV RF amplitude, which is sufficient for the proposed rebunching process. The commercial availability of such cavities is encouraging and has inspired the baseline parameters for the simulation studies below. However, in a real implementation, the frequency of the rebunching cavity must be a harmonic of the initial RF frequency \footnote{During the rebunching procedure described in the following subsections, particles leak into neighboring 53.1 MHz RF buckets. When rebunched at a frequency that is not a harmonic of 53.1 MHz, these particles increase the average time-width RMS far beyond 100-200~\psec{}.}.  Thus, we have selected the 10th harmonic, or 531 MHz.  This implies that the cavity dimensions will be slightly smaller than those of the Cornell B-cell cavity, but it is likely the same cryostat design could be used.

\subsection{Example of a rebunching procedure in the Fermilab Main Injector}
\label{simulation}

A multi-particle simulation has been performed to study the following properties of a rebunching procedure: the ramp-down/ramp-up functions for rebunching at a higher frequency, the final RMS time widths of each 531 MHz bunch, and the total additional time needed to rebunch. This section will give an overview of the first results without much procedure tuning. 

\begin{figure}[!h]
	\centering
        \includegraphics[width=8.6cm]{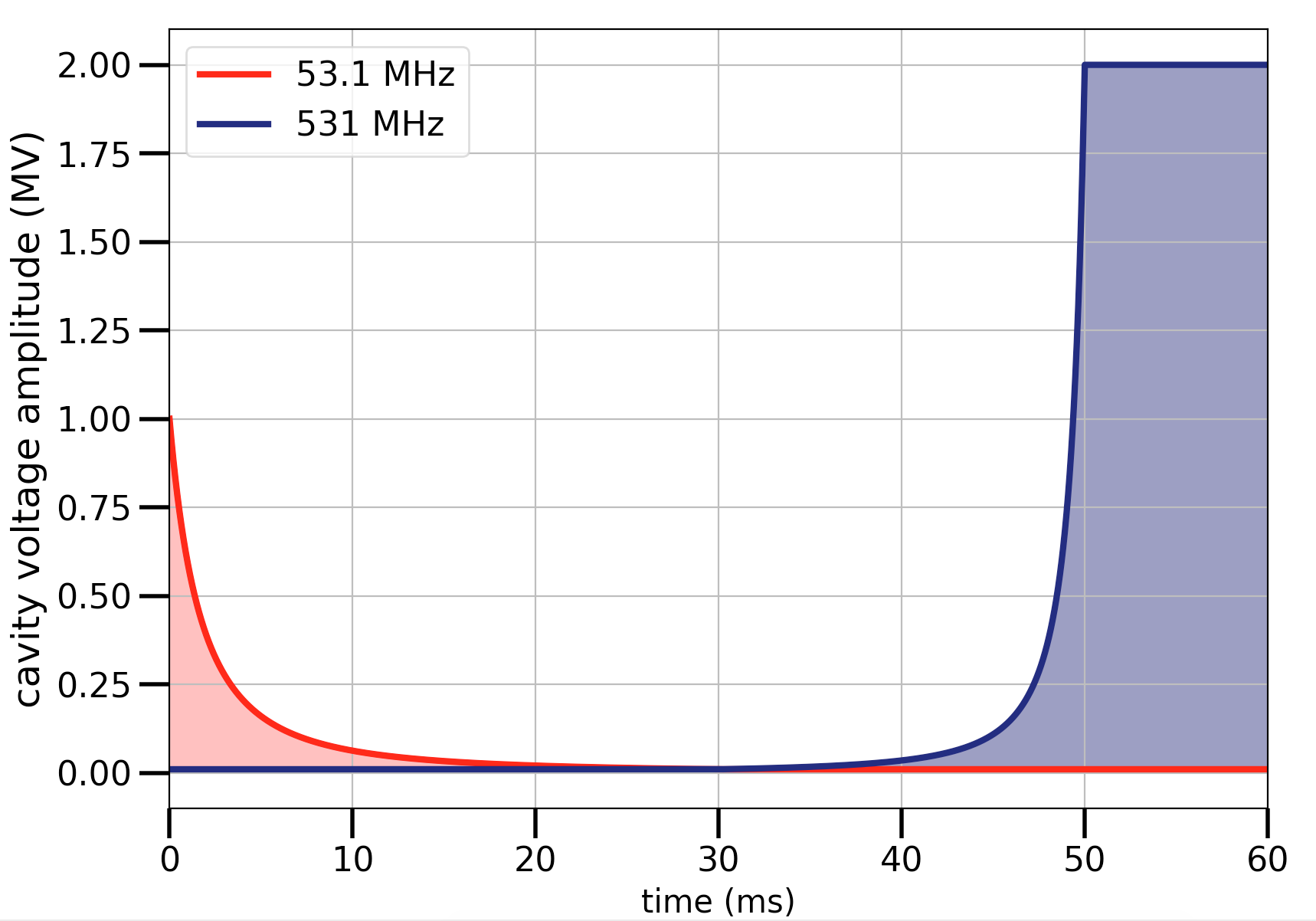}
	 
	\caption{An example of the time-evolution of the applied voltages to the 53.1 MHz
          cavities (red) and the 531 MHz cavity (blue). Both are adiabatic voltage
          ramp-up/ramp-down functions~\cite{adiabatic_capture}. A variety of time-scales were considered for the ramp-down and ramp-up rate, as well as the total time in which both cavities are at 0 V.}
		\label{fig:transition_voltages}
\end{figure}

\begin{figure}[!h]
	\centering
        \includegraphics[width=8.6cm]{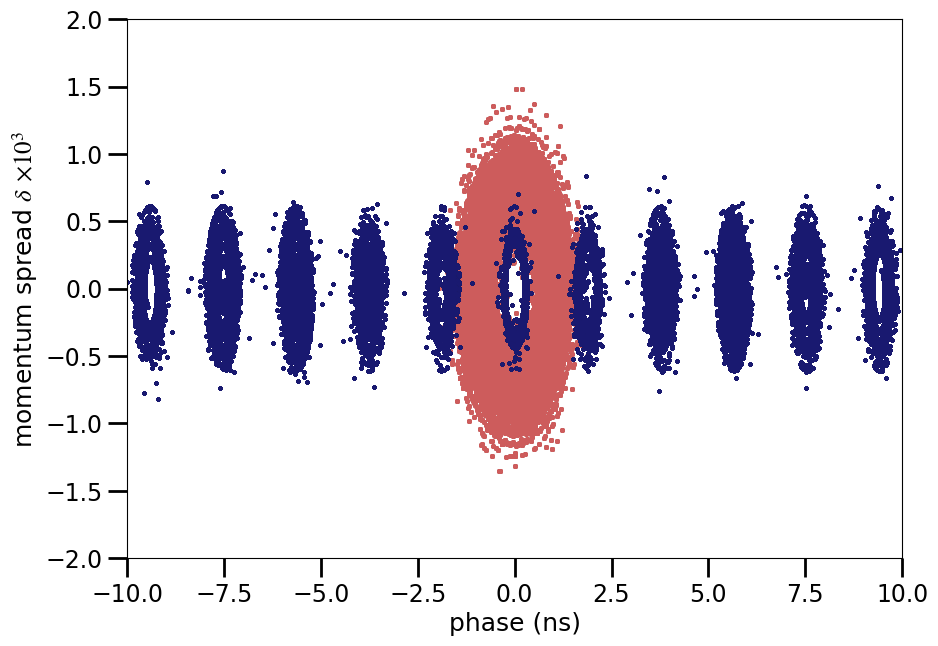}
	 
	\caption{Distribution of protons in momentum spread vs. RF phase for initial 53.1 MHz (red) and final 531 MHz (blue) states. The initial state proton bunch is taken from the main injector simulation at flat top provided by reference \cite{Ainsworth:IPAC2017-WEPVA039} and shown in Figure \ref{fig:MI_distributions}. This distribution was re-sampled to include $5\times10^6$ protons per initial 53.1 MHz bunch.}
		\label{fig:bunch_distributions}
\end{figure}

\begin{figure*}[!htb]
	\centering
        \includegraphics[width=0.9\textwidth]{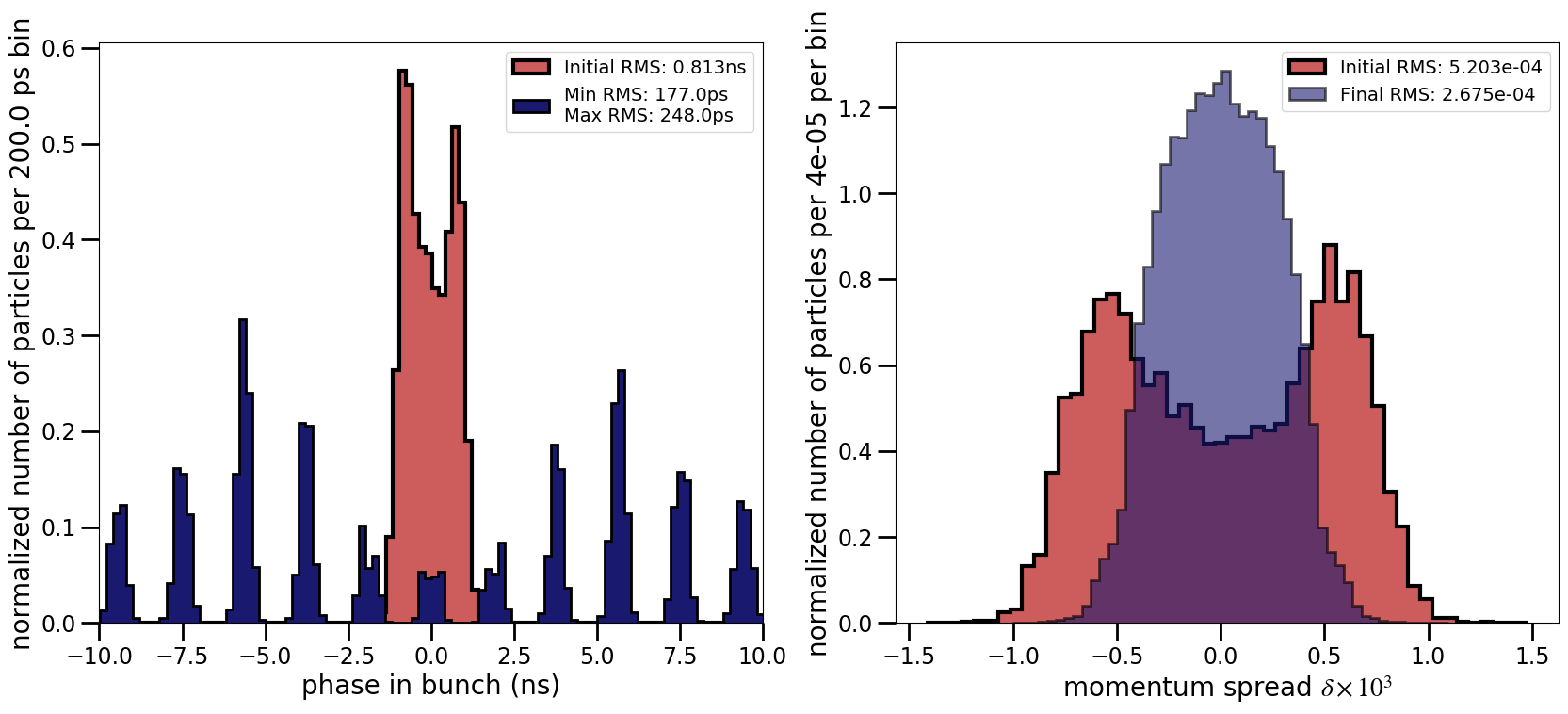}
	 
	\caption{Projection of the initial (53.1 MHz, red) and final (531 MHz, blue) bunches onto the phase space variables $\phi$ and $\delta$ using the proton bunch initial conditions from \ref{fig:MI_distributions}. The simulation includes 20 neighboring 53.1 MHz RF buckets. The resulting width of the 531 MHz bunches is affected by inter-bucket cross-talk. The number of protons contained in each 531 MHz bunch varies from 15\% to 5\% of the number of protons in the initial 53.1 MHz bunch.}
		\label{fig:bunch_projections}
\end{figure*}

\subsubsection{Simulation details}

In the simulations, each particle was tracked independently, using the following one-turn synchrotron map with synchronous phase $\phi_s = 0$\cite{SYLee}:
\begin{equation}\label{eq:delta}
\delta_{n+1} = \delta_n +  \frac{e V_{rf}}{\beta ^2 E}\sin(\phi_n) 
\end{equation}
\begin{equation}
\phi_{n+1} = \phi_n + 2\pi h \eta \delta_{n+1} 
\end{equation}
where $n$ indexes the number of turns, $E$ is the energy of the particle, $\beta$ is the usual relativistic velocity, $\eta \approx 2.4 \times 10^{-3}$ is the MI phase slip factor, and $h=588$ is the RF harmonic number.  Tracking in the simulations was performed with $2 \times 10^5$ particles.

For two RF voltage functions with frequencies $f_{1}$ and $f_{2}$
superposed, Equation \ref{eq:delta} becomes

\begin{equation}
\delta_{n+1} = \delta_n +  \frac{1}{\beta ^2 E}\big ( eV_{1}\sin(\phi_n) + eV_{2}\sin(\frac{f_2}{f_1}\phi_n) \big ) 
\end{equation}
where the ratio of two frequencies, $f_2/f_1$, is equal to 10 in this study.  The phase variable, $\phi _n$, is the phase of the particle after $n$ turns relative to the lower frequency RF voltage, 53.1 MHz. 

\subsubsection{Voltage transition functions}

The main strategy for manipulating longitudinal phase space adiabatically, i.e. without changing the emittance, is to adjust RF cavity voltages on time scales that are much longer than the synchrotron frequency at any given time.  In practice, it is difficult to achieve this condition because the synchrotron frequency tends to zero as the RF voltage is lowered. The Fermilab g-2 experiment has explored functions for adiabatic voltage changes of cavities that preserve longitudinal emittance \cite{adiabatic_capture}. If $A$ is the area of an RF bucket, and the bucket has a square shape, then the following functions can be derived from the adiabatic condition $\omega_s \gg \frac{1}{A} \frac{\mathrm{d}A}{\mathrm{d}t}$

\begin{equation}
V(t) = \frac{V_0}{[1 - (1 - \sqrt{V_0/V_1})t/t_1]^2}
\end{equation}
for ramping voltages up ($V_1 > V_0$), and 

\begin{equation}
V(t) = \frac{V_0}{[1 + (\sqrt{V_0/V_1} - 1)t/t_1]^2}
\end{equation}
for ramping voltages down. The essential parameter is $t_1$, which is the ramping time interval (10-30 ms in this study). These functions will be referred to as
\textit{adiabatic} voltage functions.

An example of a rebunching recipe using these functions is shown in Figure
\ref{fig:transition_voltages}. The end point voltage of 2 MV for the
higher frequency cavity is set by the specifications of the existing 500 MHz superconducting RF cavities \cite{nsls-cavity}. The starting voltage of 1 MV for the 53.1 MHz structure is set by the operating parameters of the Fermilab Main Injector. Timescales of $<$60 ms for the complete rebunching procedure were set to minimize the increase in the MI cycle time.

\subsubsection{Results of simulations}

An example of initial conditions and final rebunched particle longitudinal positions is shown in Figure
\ref{fig:bunch_distributions}. A projection of the longitudinal time variable and the momentum spread $\delta$ are shown in Figure \ref{fig:bunch_projections}.

The simulation shows that 150-300~\psec{} time-width RMS final bunches were achieved with little optimization of voltage functions. These particle distributions were fed into the neutrino simulations found in Section \ref{results}.

The following affects have been observed in the simulation:
\begin{itemize}
    \item The objective of the ramp-down of the 53.1 MHz voltage amplitude is to create a thin band in $(\phi, \delta)$ phase-space where the spread in $\delta$ is very small and subsequently the spread in $\phi$ is very large. Smaller $\Delta \delta$ will lead to shorter bunches after the higher frequency cavity voltage reaches 2 MV. However, because of the large $\Delta \phi$, there will be larger cross-talk between two neighboring 53.1 MHz buckets. 
    \item The final 531 MHz bunch RMS varies based on the position of the bunch relative to the original 53.1 MHz RF structure. In particular, central 531 MHz bunches are shortest and those at the edge (10~\nsec{} away) are wider. Particles with higher $\delta$ values will leak further into neighboring 53.1 MHz buckets. When these particles are rebunched, they reside closer to the separatrix of the higher harmonic RF structure and thus contribute to higher time-width RMS. This effect especially increases the average RMS of bunches when the full set of 504 MI buckets are overlayed.
    \item Kinks and non-adiabatic changes in the voltage recipes, such as cavities turning off or on, may cause unwanted longitudinal harmonics in the final 531 MHz bunches that leads to an increase in RMS width.
    \item By using more than one superconducting RF cavity, one can reduce the final bunch length by adding additional RF voltage. The momentum acceptance of the Main Injector can tolerate higher momentum spreads and, consequently, shorter bunches.
\end{itemize}


%
%

\section{Results: Neutrino Spectra from a Simulation of the Rebunched Fermilab Beam.}
\label{results}

 \begin{figure*}[!htb]
	\centering
        \includegraphics[width=0.9\textwidth]{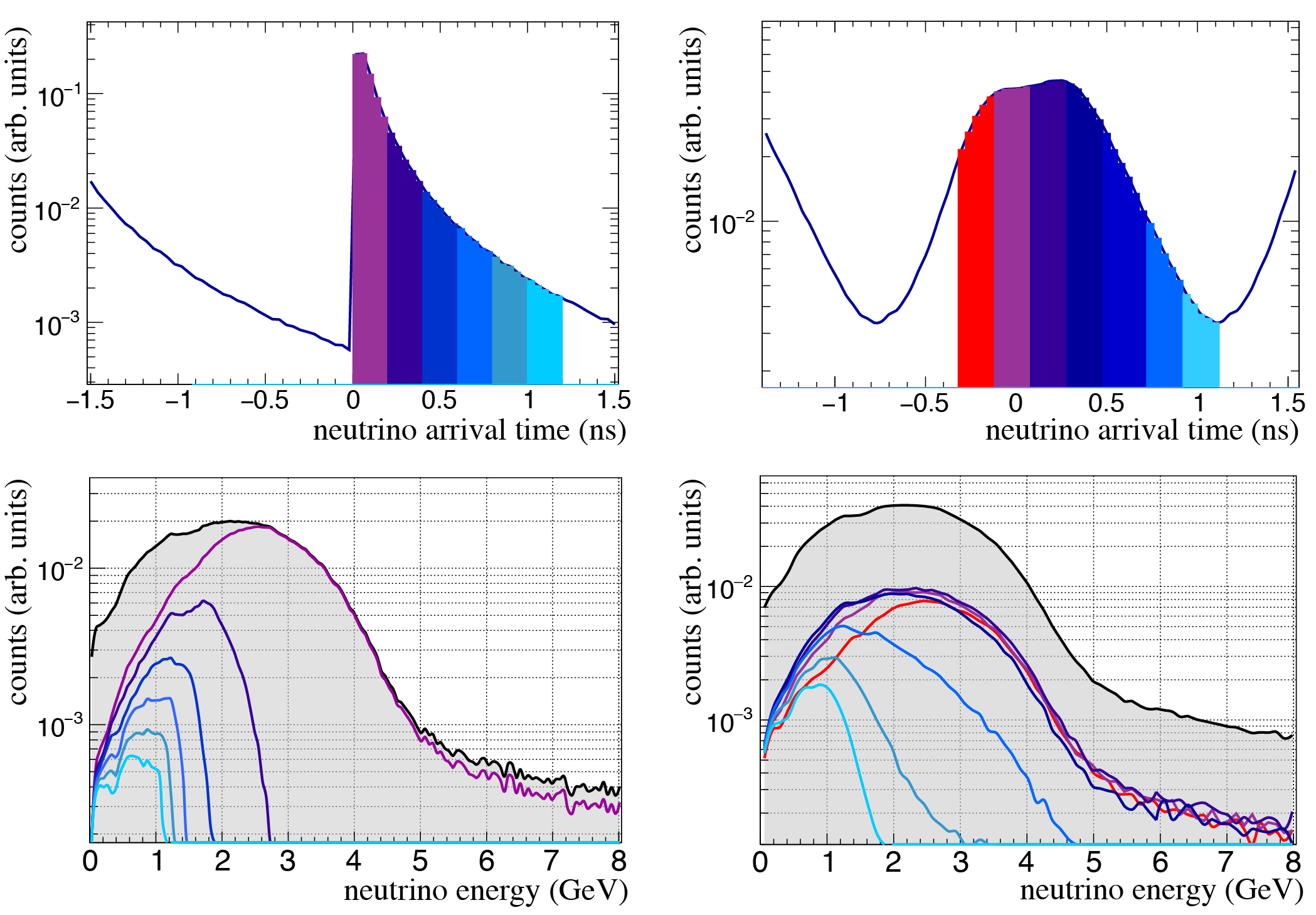} 
	\caption{The plots on the top row show the relative time-of-arrival of all neutrinos at the far detector for the zero bunch width and perfect detector timing (left) and 250~\psec{} bunch width and 100~\psec{} detector timing in 200~\psec{} bins (right). Time cut ranges that produce the fluxes on the bottom row (red, majenta, indigo, blue, light blue, cyan, teal) are shown as shaded regions in the time-of-arrival plots on the top row. All plots include pile-up affects from neighboring 531 MHz bunches. The plots on the bottom row show the simulated LBNF neutrino energy distribution (outer envelope), overlaid with the fluxes corresponding to increasingly later binned time-cuts relative to the bunch arrival time. The bins are 200~\psec{} wide in both cases. Both plots are in Forward Horn Current mode, as Reverse Horn Current versions look identical.}
		\label{fig:timeslicesFHC}
\end{figure*}

To examine a realistic scenario for the stroboscopic approach, the neutrino transit times from a Monte Carlo of the DUNE flux~\cite{Alion:2016uaj} were distributed according to the time structure of the simulated rebunched beam (simulated bunch described in Sec~\ref{RF}). An additional Gaussian smearing of 100 psec was applied as a plausible future resolution for measuring the time of neutrino interactions in the detector. These are time resolutions that may be achieved today in water cherenkov detectors~\cite{ANNIE, Theia}. There is a need for new ideas and further studies that would allow large liquid argon detectors to achieve O(100)~\psec{} resolution on neutrino interactions. 

These simulations were generated for both Forward Horn Current (FHC) and Reverse Horn Current (RHC) horn configurations. As described at the end of Sec~\ref{RF}, 531 MHz bunches near the center of the 53.1 MHz RF bucket have a smaller time-RMS than bunches near the edge, 10~\nsec{} away. In this section, we show the results of using the worst case 531 MHz bunch, with 250~\psec{} RMS, as input to the stroboscopic energy discrimination.

The plots in Fig~\ref{fig:timeslicesFHC} show the energy spectra for the FHC configuration of the LBNF beam (bottom) and the time of arrival of neutrinos at the detector (top). The arrival times of neutrinos were binned in 200~\psec{} wide bins. The neutrinos that are contained in these time-bins are plotted as individual curves in the flux/energy plots (bottom). The left column shows the ideal case where 531 MHz bunches have zero width and the neutrino detector has perfect timing resolution on the neutrino arrival time. The right column shows how both distributions smear when 250~\psec{} bunch RMS bunches are used and 100~\psec{} timing resolution is assumed at the detector. 

Because the 531 MHz bunches are separated by only 1.88~\nsec{}, high energy neutrinos that arrive early relative to the proton bunch will leak into the low energy neutrino time-bins. Energy discrimination is still possible; however, this ``pile-up'' effect adds additional degradation of energy discrimination at the low neutrino energies. With a zero RMS bunch width proton bunch, 1.27\% of neutrinos overlap with neutrinos from the following bunch. With the 250~\psec{} RMS proton bunches and 100~\psec{} detector timing resolution, 3.0\% of neutrinos from the previous bunch are overlapping with 3$\sigma$ of the neutrinos in the next bunch. 

 Even with smearing effects from finite timing resolution and the structure of the bunches, it is possible to use timing cuts to select a subset of the flux peaked in the energy range of the LBNF second oscillation maximum~\cite{DUNE2ndMax} and to suppress backgrounds from higher energy neutrinos such as neutral current resonant pion events~\cite{PionProductionThreshold}.

 \begin{figure*}[!htb]
	\begin{center}
        \includegraphics[width=0.9\textwidth]{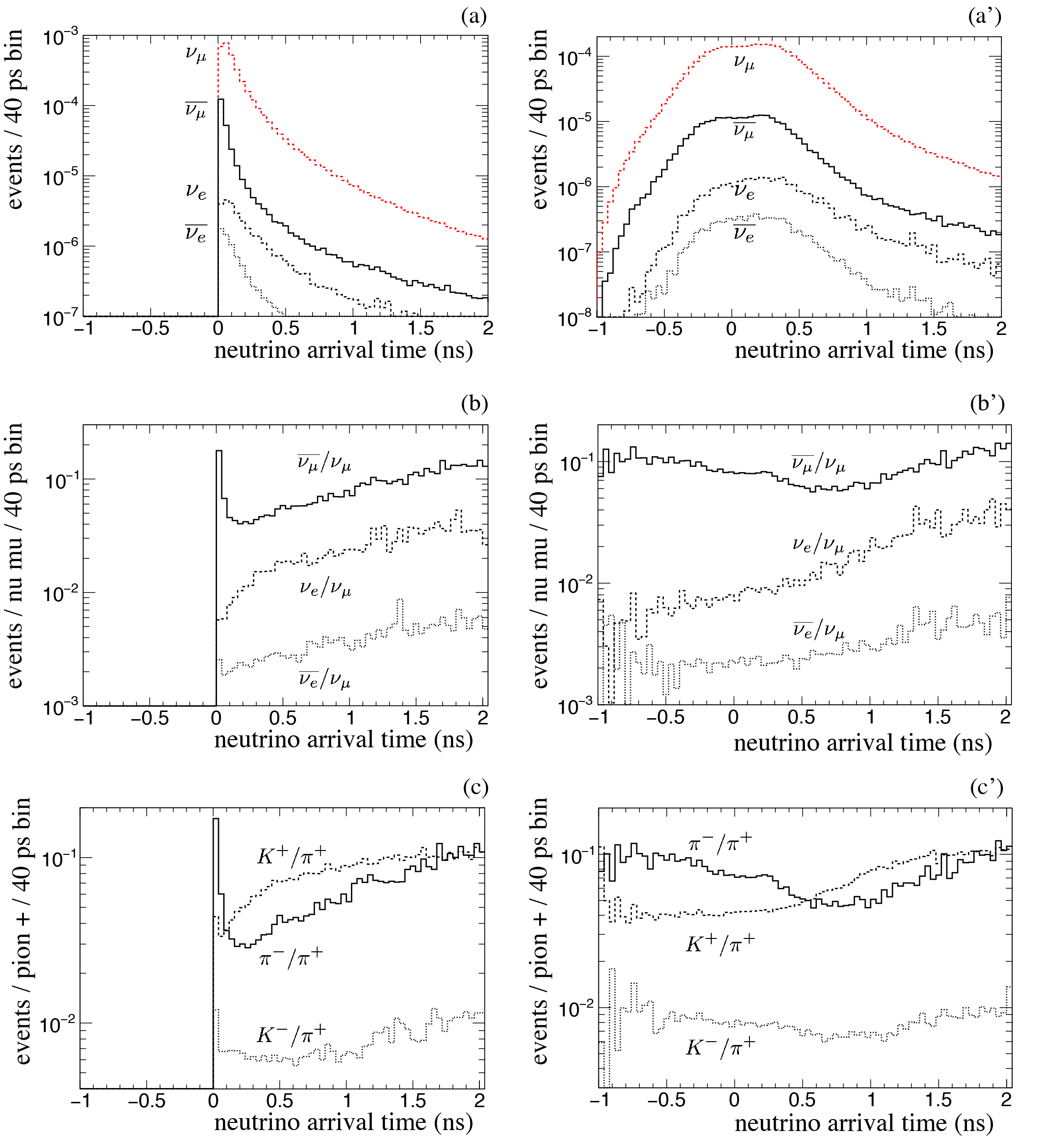}
	\end{center}
	\caption{Simulated arrival time distributions of neutrino flavors and neutrino parent hadrons. (a) time of arrival distribution for the most common neutrino flavors given a zero RMS proton bunch width and perfect detector timing. (b) neutrino flavor arrival times normalized to the most common flavor, $\nu _{\mu}$. (c) neutrino parent arrival times normalized to the most common parent, $\pi ^{+}$. All plots in the right hand column (a', b', c') include the effects of 250~\psec{} bunch widths, 100~\psec{} detector timing resolution, and pile-up from neighboring bunches. All plots are Forward Horn Current mode. }
		\label{fig:neutrinocompositionintimesmeared}
\end{figure*}

 Studying the time evolution of the flux may also open the possibility of studying changes in the flavor, sign, or parent hadron composition of the beam in time. The top row of figure~\ref{fig:neutrinocompositionintimesmeared} shows the neutrino flavor composition as a function of neutrino arrival time relative to the proton bunch. Smearing is applied using 250~\psec{} bunch widths and 100~\psec{} detector resolution. The parent hadron and flavor contributions are shown in the lower two rows normalized to the primary parent or flavor ($\pi ^+$ and $\nu _{\mu}$ respectively). 

The results in this section were derived for the far detector but apply equally to the near detector. For the near detector the stroboscopic approach can be applied together with prismatic measurements~\cite{Nuprism, DUNEprism} that sample multiple off-axis angles. In combination, the two techniques provide additional constraints on flux and cross-section uncertainties.


%
%
\section{Conclusions}
\label{conclusions}

A single superconducting cavity at the 10th RF frequency harmonic of the current 53 MHz RF structure in the Fermilab Main Injector (MI) would allow using precision time measurements to statistically discriminate the energy and possibly the family and flavor of neutrino events in on-axis detectors. For narrow proton bunches, the time of the neutrino interaction in the near or far neutrino detector relative to the production of the parent hadron in the target is weighted towards lower energy for later events and higher energy for earlier events. Selecting on time bins (the `stroboscopic' technique) relative to the proton bunch thus performs a similar sculpting of the spectrum for an on-axis geometry as going off-axis without losing the on-axis flux. The relative fractions of $e$, $\mu$, and $\tau$ neutrinos also change with time bin and with polarity. Timing could be used to search for the prompt production of new particles such as light neutral leptons~\cite{shrock} or light dark matter~\cite{DUNEprismLDM, deGouvea2019}. The stroboscopic approach provides additional constraints to the flux and cross-section uncertainties that will dominate future oscillation measurements.

The technique relies on the implementation of  precision time measurements at the neutrino detector relative to those at the proton target. The simulations from the present study indicate that timing resolution and proton bunch RMS should be optimized so that they are comparable in magnitude. Precision timing with respect to the beam is in reach of current technology for near detectors. For far detection, this work will hopefully motivate further advances in precision timing transfer, a renewed interest in implementing fast timing capability in LAr-TPC detectors~\cite{Foreman:2019bjd, Kryczynski:2016opr, Szelc1}, and the development of higher precision timing photodetector systems in water- and scintillator-based detectors~\cite{ANNIE, Theia}. Precision timing using pico-second photo-detectors and improved event vertex reconstruction algorithms~\cite{Aberle2014, ANNIEreco, Wonsak2018, Sakai_thesis} will be necessary.

We provide as an example a specific implementation of rebunching the Main Injector  53 MHz at flat-top with hardware parameters that are feasible with present commercial technology ~\cite{nsls-cavity}. A simulation was made of the 53.1 MHz ramp-down and 531 MHz ramp-up in order to determine the final RMS bunch width and the additional time in the cycle to rebunch. The results of the simulation show that adiabatic longitudinal manipulations can be performed on a time-scale that does not significantly affect the total protons on target and that bunches on the order of 200~\psec{} RMS can be made without extensive procedural tuning. More detailed studies can improve upon the final RMS and also implement more realistic experimental parameters.

\section{Acknowledgements}
\label{thanks}
This manuscript has been co-authored by The Fermi Research Alliance, LLC
under Contract No.  DE-AC02-07CH11359 with the U.S. Department of
Energy, Office of Science, Office of High Energy Physics.
The work at the University of Chicago 
was supported by DOE contract DE-SC-0008172 and NSF Contract No. PHY-1707981. The work at 
Iowa State University) was performed under contract DE-SC0017946. We thank Robert
Tschirhart for a leading comment at the March 17, 2018 Fermilab-Chicago
meeting on Psec Timing~\cite{ChicagoTiming}. We also thank Robert Ainsworth for providing up-to-date information about the MI beam properties and Phil Adamson for early discussions regarding precision beam timing. A special thanks Laura Fields and Zarko Pavlovic for all of their help in accessing and parsing the DUNE flux files.

\newpage 
\bibliography{main}

\end{document}